\documentstyle[preprint,aps]{revtex}
\newcommand{\beq}{\begin{equation}}
\newcommand{\eeq}{\end{equation}}
\newcommand{\beqa}{\begin{eqnarray}}
\newcommand{\eeqa}{\end{eqnarray}}
\newcommand{\ba}{\begin{array}}
\newcommand{\ea}{\end{array}}

\begin{document}
\draft


\widetext 
\title{Thermodynamics of Multi-Component Fermi Vapors} 
\author{L. Salasnich$^{(*)(+)}$, B. Pozzi$^{(*)(+)}$, A. Parola$^{(*)(++)}$ 
and L. Reatto$^{(*)(+)}$} 
\address{$^{(*)}$ Istituto Nazionale per la Fisica della Materia, 
Unit\`a di Milano, \\
Via Celoria 16, 20133 Milano, Italy \\
$^{(+)}$ Dipartimento di Fisica, Universit\`a di Milano, \\
Via Celoria 16, 20133 Milano, Italy \\ 
$^{(++)}$ Dipartimento di Scienze Fisiche, Universit\`a dell'Insubria, \\ 
Via Lucini 3, 23100 Como, Italy}
\maketitle

\begin{abstract} 
We study the thermodynamical properties of Fermi vapors  
confined in a harmonic external potential.  
In the case of the ideal Fermi gas,  
we compare exact density profiles  
with their semiclassical approximation  
in the conditions of recent experiments. 
Then, we consider the phase-separation of a  
multi-component Fermi vapor. In particular,  
we analyze the phase-separation as a function of temperature,  
number of particles and scattering length.  
Finally, we discuss the effect of rotation  
on the stability and thermodynamics of the trapped vapors.  
\end{abstract} 
 
\vskip 0.5cm 
 
\pacs{PACS Numbers: 05.30-d, 05.30.Fk}  
 
 
\narrowtext 
 
\section{Introduction} 
 
Recent experiments with dilute vapors of  
alkali-metal atoms in magnetic or  
magneto-optical traps at very low temperatures have renewed  
the theoretical study of Bose and Fermi gases.  
For atomic gases, the Bose-Einstein condensation  
has been achieved in 1995 [1-3] and the Fermi  
quantum degeneracy in 1999 [4,5].  
\par 
The s-wave scattering between Fermions in the same hyperfine state  
is inhibited due the Pauli principle.  
It follows that at low temperature the dilute Fermi gas,  
in a fixed hyperfine state, is practically ideal.  
Nevertheless, the effect of interaction could be very effective  
for a Fermi vapor with two or more hyperfine states (components).  
In the recent experiment with dilute $^{40}$K Fermions [4,5],  
to favor the evaporative cooling,  
a $^{40}$K Fermi vapor in two hyperfine states  
($|9/2,9/2\rangle$ and $|9/2,7/2\rangle$) is used.  
When the system is below the Fermi temperature,  
one hyperfine component is removed and  
it remains a trapped quasi-ideal degenerate Fermi gas.  
\par  
In this paper we analyze the thermodynamical properties  
of a many-component interacting Fermi vapor by using  
a harmonic potential that models the trap of recent experiment  
with $^{40}$K [4,5]. These trapped Fermi gases  
are quite interesting because the quantum degeneracy shows up not  
only in momentum space, as in uniform systems, but also  
in coordinate space. First, we investigate the accuracy of the  
semiclassical approximation for the study of the  
properties of the Fermi gas in the conditions of $^{40}$K experiment. 
For the ideal Fermi gas we discuss the finite-temperature 
spatial and momentum distribution and derive, in the large impulse 
approximation, the dynamic structure factor. 
Then, we consider Fermi vapors with many components.  
Due to the nonzero interaction between different hyperfine states,  
one can obtain the phase-separation of Fermi components  
by varying temperature, number of particles and  
scattering length. Finally, we discuss the effect of an imposed  
rotation around a fixed axis on stability and thermodynamics  
of the trapped Fermi vapors.  
 
\section{Ideal Fermi Gas} 
 
For an ideal Fermi gas at thermal equilibrium,  
the average number of particles in the single-particle  
state $|\alpha\rangle$ with energy $\epsilon_{\alpha}$ is given by  
\beq  
n_{\alpha} = {1\over e^{\beta(\epsilon_{\alpha}-\mu)} + 1}  \; ,  
\label{eq1}  
\eeq  
where $\mu$ is the chemical potential and $\beta=1/(kT)$  
with $k$ the Boltzmann constant and $T$ the temperature.  
The average number $N=\sum_{\alpha} n_{\alpha}$  
of particles of the system fixes the chemical potential [6].  
In the case of a harmonic potential  
$U({\bf r}) = (m/2)(\omega_1^2 x_1^2 + \omega_2^2 x_2^2  
+\omega_3^2 x_3^2)$ one finds the exact density profile  
by using the Eq. (\ref{eq1}) and Hermite polynomials,  
from which one obtains the eigenfunctions $\phi_{n_1n_2n_3}({\bf r})$ 
of the harmonic oscillator  
\beq 
n({\bf r})= \sum_{n_1n_2n_3=0}^{\infty}  
{|\phi_{n_1n_2n_3}({\bf r})|^2  
\over e^{\beta\hbar(\omega_1(n_1+1/2)+\omega_2(n_2+1/2)  
+\omega_3(n_3+1/2)-\mu)} + 1} \; .  
\label{eq2} 
\eeq  
\par 
The exact density profile can be compared with the semiclassical one. 
The semiclassical approximation means that,  
instead of $\epsilon_{\alpha}$,  
one uses the classical single-particle phase-space energy 
$\epsilon({\bf r},{\bf p})={{\bf p}^2\over 2m} + U({\bf r})$. 
In this way one obtains the single-particle phase-space distribution 
\beq 
n({\bf r},{\bf p}) = 
{1\over e^{\beta(\epsilon({\bf r},{\bf p})-\mu)} + 1}  \; . 
\label{eq3}
\eeq 
At zero temperature, i.e., in the limit $\beta\to \infty$ where 
$\mu= E_F$ (the Fermi energy), one has $n({\bf r},{\bf p}) = \Theta 
\left(E_F - \epsilon({\bf r},{\bf p}) \right)$, 
where $\Theta(x)$ is the Heaviside step function. 
At temperature well below the Fermi temperature 
$T_F=E_F/k$, the Fermions begin to fill 
the lowest available single-particle states 
in accordance with the Pauli exclusion Principle 
(Fermi quantum degeneracy) [6]. 
\par 
Because of the Heisenberg principle, 
the quantum elementary volume of the single-particle phase-space is 
given by $(2\pi\hbar)^3$, where $\hbar$ is the Planck constant. 
It follows that the average number $N$ of particles can be written as 
\beq
N= \int {d^3{\bf r} d^3{\bf p} \over (2\pi \hbar )^3}
\; n({\bf r},{\bf p}) =
\int d^3{\bf r} \; n({\bf r}) = 
\int d^3{\bf p} \; \tilde{n}({\bf p}) \; ,
\label{eq4}
\eeq
where $n({\bf r})$ is the spatial distribution and 
$\tilde{n}({\bf p})$ is the momentum distribution. 
The finite temperature spatial distribution is given by  
\beq 
n({\bf r})={1\over \lambda^3} f_{3/2} 
\left(e^{\beta(\mu -U({\bf r}))}\right) \; , 
\label{eq5} 
\eeq 
where $\lambda = (2\pi \hbar^2\beta /m)^{1/2}$ is the  
thermal length and  
\beq 
f_{n}(z) = {1\over \Gamma(n)} \int_0^{\infty} dx  
{x^{n-1}\over z^{-1}e^{x}+1} \; , 
\label{eq6} 
\eeq 
with $\Gamma$ the factorial function. 
In the limit of zero temperature, with $\mu=E_F$ the Fermi energy, 
the spatial distribution gives the Thomas-Fermi approximation 
$n({\bf r})=(2m)^{3/2}/(6\pi^2\hbar^3) 
\left(E_F- U({\bf r})\right)^{3/2}  
\Theta\left(E_F- U({\bf r})\right)$, 
where $\Theta$ is the Heaviside step function [7-8]. 
The Eq. (\ref{eq5}) is the generalization of well-known formula  
for an ideal homogenous Fermi gas in a box that is 
exact in the thermodynamical limit [6].  
Note that for $|z|<1$ one has $f_{n}(z)=  
\sum_{i=1}^{\infty} (-1)^{i+1} z^i/i^n$.  
Moreover, by using $g_n(z)=-f_n(-z)$ instead of  
$f_n(z)$, one finds the spatial distribution  
of the ideal Bose gas in external potential.  
\par  
A comparison between exact and semiclassical results has been  
recently performed at zero [9] and finite temperature [10].  
It has been shown that the semiclassical approximation  
is good for $(kT/\hbar \omega_H)>>1$,  
where $\omega_H=(\omega_1\omega_2\omega_3)^{1/3}$, or,  
at a fixed temperature, for a large number $N$ of trapped particles.  
Our calculations confirm this prediction.  
In fact, we find that when $(kT/\hbar \omega_H)>1$ there are  
no appreciable deviations between exact and semiclassical results.  
Instead, when $(kT/\hbar \omega_H)<1$ some differences are  
observable, in particular for "magic" numbers of particles  
($N=1,4,10,20,35,56,84,...$) that correspond to a complete shell 
occupation of single-particle energy levels (see also [10]). 
The differences are reduced by increasing $N$ 
showing that semiclassical approximation provides an excellent 
representation of Fermi distribution for a wide range of 
parameters. In Figure 1 we plot the exact density profile, obtained 
by a direct calculation of Eq. (\ref{eq2}), and 
that obtained with the semiclassical function (\ref{eq5}) 
for $(kT/\hbar \omega_H)=10^{-3}$ and three "magic" numbers. 
In correspondence of the "magic numbers", the spatial density 
profile shows local maxima, which suggest a spatial shell structure. 
The magic numbers are particularly stable; 
in fact, for small variations of the chemical potential $\mu$ 
the magic number $N$ remains unchanged. 
The shell structure in the density profile 
is washed out by increasing the number of particles 
and is completely absent in the semiclassical approximation. 
\par 
As shown in the experiment [4], for the quasi-ideal one-component  
Fermi gas of $^{40}$K atoms confined in a cigar-shaped trap  
($\omega_1=\omega_2=860.80$ Hz, 
$\omega_3=122.52$ Hz), the semiclassical formulas of an ideal 
Fermi gas in a harmonic trap agree with the available 
experimental data within the statistical uncertainty. 
The accuracy of the semiclassical predictions are 
not surprising because the experimental conditions are 
$kT/(\hbar \omega_H)\simeq 10^2$-$10^3$ and $N\simeq 10^6$. 
For the sake of completeness, in Table 1 we show some relevant 
quantities of the Fermi gas at different temperatures: 
the energy $E=3/(2\beta \lambda^3)\int d^3{\bf r} 
f_{5/2}\left(e^{\beta \left(\mu -U({\bf r})\right)}\right) 
+\int d^3{\bf r} U({\bf r}) n({\bf r})$, 
the density at the origin $n({\bf 0})$ 
and the widths $\langle\rho^2\rangle^{1/2}$ 
and $\langle z^2\rangle^{1/2}$ of the Fermi cloud in the 
radial and axial directions, respectively. 
\par 
In addition to the spatial distribution $n({\bf r})$, 
also the momentum distribution $\tilde{n}({\bf p})$ gives an useful 
signature of the quantum degeneracy of the Fermi gas. 
In the semiclassical approximation, 
the finite temperature momentum distribution 
of an ideal Fermi gas in a harmonic trap 
is calculated from Eq. (\ref{eq4}) and reads 
\beq 
\tilde{n}({\bf p})={1\over (m\omega_H\lambda)^3} 
f_{3/2}\left(e^{\beta(\mu- {{\bf p}^2\over 2m} )}\right) \; . 
\label{eq7}
\eeq 
At zero temperature, one recovers the formula 
found by Butts and Rokhsar [7], namely 
$\tilde{n}({\bf p})=1/(6\pi^2\hbar^3)2^{3/2}/(m\omega_H^2)^{3/2} 
(E_F-{\bf p}^2/2m)^{3/2}\Theta(E_F -{\bf p}^2/2m)$. 
Despite the spatial anisotropy of the trap, the momentum distribution 
of the Fermi gas is isotropic when the semiclassical approximation 
is used. Moreover, it is not difficult to show that,  
both in the exact and semiclassical cases, 
the momentum distribution $\tilde{n}$ 
can be found from the spatial one $n$ by the simple relation 
$\tilde{n}\left(p_1,p_2,p_3\right)=
n\left(p_1/(m\omega_1),p_2/(m\omega_2), 
p_3/(m\omega_3)\right)/(m\omega_H)^3$, 
where ${\bf p}=(p_1,p_2,p_3)$. In general, the 
exact momentum distribution is not isotropic and it presents 
the same shell structures of the spatial one when $N$ is close 
to a magic number. Finally, the Eq. (\ref{eq7}) can be used to calculate 
the dynamic structure factor $S({\bf q},E)$ of the Fermi gas. 
In the large impulse approximation [11] one finds 
the remarkable result 
\beq 
S({\bf q},E) = {2\pi m^2\over q\beta} 
{1\over (m\omega_H\lambda )^3} 
f_{5/2}\left(e^{\beta\left(\mu- {(E-{\bf q}^2/2m)^2 m 
\over 2{\bf q}^2} \right)} \right) \; ,
\label{eq8}
\eeq 
where ${\bf q}$ and $E$ are the momentum and energy transferred by 
the probe to the sample. Recently, the dynamic structure factor of 
a trapped Bose-Einstein condensate has been measured using 
two-photon optical Bragg spectroscopy with a time of flight technique 
by which the number of optically excited atoms can be counted [12]. 
 
\section{Multi-Component Fermi Vapor} 
  
The problem of a dilute Fermi vapor with $M$  
hyperfine states (components) 
can be studied by using the s-wave scattering approximation, 
the mean-field approach and semiclassical formulas. 
The spatial density profile $n_i({\bf r})$ of the $i$-th  
component of a Fermi vapor can be written as  
\beq  
n_i({\bf r})={1\over \lambda^3}  
f_{3/2}\left(e^{\beta \left(\mu_i -U({\bf r}) - \sum_{j=1}^M  
g_{ij} n_j({\bf r}) \right)}\right) \; ,  
\label{eq9} 
\eeq  
where $i=1,2,...,M$, $\mu_i$ is the chemical potential of the  
$i$-th component, and $g_{ij}=4\pi\hbar^2a_{ij}/m$, with $a_{ij}$ the  
s-wave scattering length between $i$-th and $j$-th component ($a_{ii}=0$).  
Thus, the effect of the other $M-1$ Fermi components on  
the $i$-th component is the appearance of a mean-field effective potential.  
Note that in the limit of zero temperature, one finds the equations  
used by Amoruso {\it et al} [13].  
The total energy of a Fermi vapor with $M$ components is given by  
$$ 
E=\sum_{i=1}^M {3 \over 2} {1\over \beta \lambda^3}  
\int d^3{\bf r}\; f_{5/2}\left(e^{\beta \left(\mu_i -U({\bf r})  
- \sum_{j=1}^M g_{ij} n_j({\bf r}) \right)}\right)  
$$ 
\beq 
+ \sum_{i=1}^M \int d^3{\bf r}\; U({\bf r}) n_i({\bf r})  
+ \sum_{i<j=1}^M \int d^3{\bf r}\; g_{ij} n_i({\bf r}) n_j({\bf r}) \; , 
\label{eq10} 
\eeq  
where the first term is the kinetic energy $E_{kin}$,  
the second term is the external potential energy $E_{ext}$  
and the third term is the interaction energy $E_{int}$.  
At zero temperature the kinetic energy assumes the  
familiar Thomas-Fermi form  
$E_{kin}=\sum_{i=1}^M (6\pi^2)^{2/3} ({3\hbar^2})/(5m)  
\int d^3{\bf r} [n_i({\bf r})]^{5/3}$.  
\par 
We numerically solve the set of equations (\ref{eq9})  
with a self-consistent iterative procedure.  
Recently, we have used a similar iterative procedure  
to study $^7$Li Bose gas [14].  
If the components of the Fermi vapor  
are non-interacting then they can occupy the same spatial region.  
Instead, if the interaction is strong enough (repulsive interaction) 
or for $N$ very large one finds 
a phase-separation, i.e., the Fermi components 
stay in different spatial regions. This effect, that has been first  
discussed at $T=0$ by Amoruso {\it et al} [13], is shown 
in Figure 2, where we plot the density profiles of two components  
of the $^{40}$K Fermi vapor for different values of  
the scattering length $a$ (a=$a_{12}$)  
and $N=0.5\cdot 10^7$ atoms in each component.  
\par 
When two components have the same number of particles, the onset of  
phase-separation is also an example of spontaneous symmetry breaking.  
In fact, if the chemical potentials $\mu_i$ of the two components  
are equal, Eq. (\ref{eq9}) always admits a symmetric solution  
$n_1({\bf r})=n_2({\bf r})$. However, for particle number $N$  
larger than a threshold $N_c$ the solution bifurcates and a pair of  
symmetry breaking solutions appears. Just beyond threshold the asymptotic  
solutions begin to differ from the symmetric one in  
a neighborhood of the origin ${\bf r}={\bf 0}$, i.e. at the point of higher  
density. Therefore, an analytic formula for the critical chemical  
potential $\mu_c$ can be obtained by standard bifurcation analysis  
of Eq. (\ref{eq9}) at the origin, which gives  
\beq  
{g\beta \over \lambda^3} e^{\beta\left(\mu - g n({\bf 0})\right)}  
f_{3/2}'(e^{\beta\left(\mu - g n({\bf 0})\right)}) = 1 \; .  
\label{eq11} 
\eeq  
At $T=0$, by the use of the first term of the large $z$ expansion  
of $f_{3/2}(z)$ [6], one finds analytical expressions  
for the critical density $n_c({\bf 0})$  
and the critical chemical potential $\mu_c$:  
\beq 
n_c({\bf 0})={\pi\over 48 a^3} \; , \;\;\;\;  
\mu_c={5 \pi^2 \over 24} \hbar\omega_H  
\left({a_H\over a}\right)^2 \; .   
\label{eq12} 
\eeq  
These remarkably simple formulas can be very useful  
to determine the onset of phase-separation in  
future experiments. Moreover, by knowing the critical chemical potential  
one numerically finds the number of particles via Eq. (\ref{eq9}).  
\par  
At finite temperature, we numerically solve the Eq. (\ref{eq11}).  
In Figure 3 we plot the critical number of Fermions as a  
function of the temperature for different values of the  
scattering length. Figure 3 shows that, by using  
a realistic scattering length [4,5], 
one needs about $N=10^{12}$ atoms 
to get the spontaneous symmetry breaking. 
One sees that at higher temperatures  
the phase-separation appears with a larger number of particles  
or a larger scattering length.  
In the lower part of Figure 3 we plot the same critical line as a  
function of temperature for different values of the scattering  
length. The conclusion is that  
by increasing the interaction between the two components  
one can use lower number of particles  
to obtain the phase-separation.  
Such a behavior can be extracted form Eq. (\ref{eq12})  
in the case of zero temperature. Note that at finite temperature one  
can use the first two terms of the large $z$ expansion of $f_{3/2}(z)$ [6].  
Under the conditions $kT << \hbar \omega_H$  
and $\mu >> g n({\bf 0})$ one finds the following equation  
for the onset of phase-separation  
\beq 
\left( {a\over a_H} N^{1/6} \right)  
\left[ 
1 - {\pi^2\over 2^{11/3} 3^{5/3} } {1\over N^{2/3}}  
\left( {kT\over \hbar \omega_H} \right)^2  
\right]  
={\pi \over 2^{5/3} 3^{1/6} } \; .  
\label{eq13} 
\eeq  
\par 
One important point is to understand what happens when  
the two components of the Fermi vapor have a different number of atoms.  
In such a case, it is still possible to observe a phase-separation  
of the two components and their position is related  
to the ratio $N_1/N_2$, where $N_1$ and $N_2$  
are the number of Fermions in the two hyperfine states.  
In Figure 4 we plot the density profiles of the two components  
with different scattering length and ratios $N_1/N_2$.  
The component with less atoms is pushed outward but this  
effect appears only when the scattering length exceeds  
the value that produces the onset of phase separation with $N_1=N_2$.  
\par 
Phase-separation also appears in a Fermi vapor  
with three or more components.  
In Figure 5 we plot the density profiles of the $^{40}$K Fermi vapor  
with three components. In this case we numerically solve Eq. (\ref{eq9}) 
with $a_{12}=a_{13}=a_{23}=a$. 
The Figure shows that also for three components 
the spontaneous symmetry breaking and the 
phase-separation are controlled by scattering length, 
temperature and total number of particles. 
In particular, one finds that by increasing the scattering length 
at first one of the components separates from the others, 
which remain still mixed. Note that the separation begins at 
the center of the trap. As the scattering length is further 
increased, also these two components separate 
(this second phase-separation begins at the interface 
with the previously separated component) and 
one eventually sees complete phase-separation and 
the formation of $4$ or $5$ shells. 
\par 
The Eq. (\ref{eq11}) can be extended to a M-component Fermi 
vapor with the same number of particles in each component.  
The critical density $n_c({\bf 0})$ does not depend on the 
number $M$ of Fermi components and one gets the same result of 
Eq. (\ref{eq12}). Instead, the critical chemical potential reads 
$\mu_c=(2M+1)\pi^2 \hbar \omega_H (a_H/a)^2/24$. 
\par 
Nowadays it is possible to confine mixtures of Fermions 
and Bosons. If Bosons occupy the external region 
of the trap then the internal Fermi gas can increase its density 
at the origin and so favor the onset of phase-separation. 
We have verified that such effect is possible when 
two conditions are satisfied:  
1) the number of Bosons is close to that of Fermions;  
2) the Boson-Boson interaction is close to the Boson-Fermion one    
with scattering length of at least $10^3$ Bohr radii.  
Nevertheless the gain is limited: the density at the origin  
grows by $20\%$ for $N=10^7$ particles. In conclusion, 
to obtain the phase-separation with several millions of Fermions, 
the only chance is to strongly enhance the scattering length, 
for example by using the Feshbach resonances. The Feshbach 
resonances, already seen in Bose condensates [15], 
should allow to vary the interaction 
strength between atoms. One could use them to go 
smoothly from a quasi-ideal to a strongly interacting 
Fermi system. 
 
\section{Rotating Fermi Vapor} 
 
\par
The semiclassical approximation can be easily applied to a 
rotating gas. For an ideal gas, that is rotating 
with angular velocity $\Omega$ around  the $x_3$-axis, 
the classical single-particle energy is 
$\epsilon({\bf r},{\bf p})={{\bf p}^2\over 2m} +  
U({\bf r}) - \Omega (x_1p_2 - x_2 p_1)$.  
One immediately finds that the spatial distribution of an ideal  
Fermi gas is given by  
\beq 
n({\bf r})={1\over \lambda^3} f_{3/2} 
\left(e^{\beta\left(\mu -U({\bf r})+{1\over 2}m\Omega^2 \rho^2\right)} 
\right) \; ,  
\label{eq14} 
\eeq 
where $\rho=(x_1^2+x_2^2)^{1/2}$.  
Note that the effective potential of the rotating ($\Omega >0$) Fermi gas is  
$U({\bf r}) - (m/2)\Omega \rho^2$.  
If $U({\bf r}) \sim \rho^s$ with $s<2$ for $\rho\to \infty$ then  
the effective potential goes to $-\infty$ for $\rho \to \infty$  
and the spatial integral of the density profile is divergent.  
It follows that for $s<2$ the system is unstable against rotation.  
The same result applies for ideal Bose and Boltzmann gases. 
Obviously, the Fermi gas can remain confined 
near the center of the trap for a very long time 
depending on the tunneling probability. 
\par  
In the case of the harmonic potential,  
$s=2$ and the gas is stable against rotation  
until $\Omega$ is smaller than the smaller of the harmonic frequencies  
$\omega_1$ and $\omega_2$.  
The effect of the rotation is simply a shift in two  
trap frequencies:  
$\omega_1 \to (\omega_1^2 - \Omega^2)^{1/2}$ and   
$\omega_2 \to (\omega_2^2 - \Omega^2)^{1/2}$. In this way  
one gets the Fermi energy of the rotating ideal gas  
\beq 
E_F(\Omega ) = 6^{1/3} \hbar \omega_H N^{1/3}  
\left(1-{\Omega^2\over \omega_1^2}\right)^{1/6}  
\left(1-{\Omega^2\over \omega_2^2}\right)^{1/6} \; .  
\label{eq15} 
\eeq 
A similar formula has been found in [16] for  
the Bose-Einstein transition temperature of a Bose gas  
in axial-symmetric harmonic potential. 
Also the density profile of the rotating quantum gas is modified 
according to the frequency shift discussed above. 
\par 
The discussion can be extended to a Fermi vapor with 
two or more components. In particular, 
one can use Eq. (\ref{eq9}) and all the other equations 
by simply shifting the trap frequencies. Such a shift 
determines an enhancement of the characteristic length of the 
harmonic trap: $a_H \to a_H/\left[(1-{\Omega^2/\omega_1^2})^{1/12} 
(1-{\Omega^2/\omega_2^2})^{1/12}\right]$. It means that 
the rotating gas is more dilute and consequently 
the interatomic interaction is less effective. 
As shown in Eq. (\ref{eq12}) and Eq. (\ref{eq13}), 
the rotation increases the critical number 
of Fermions that are necessary to get phase-separation. 
 
\section{Conclusions}  
 
First of all we have analyzed the  
accuracy of the semiclassical approximation.  
We have found that the semiclassical approach is very  
accurate for $kT> \hbar \omega_H$  
or, at a fixed temperature, for a large number of particles.  
In particular, the semiclassical formulas are  
extremely reliable to study recent experiments 
with a trapped quasi-ideal $^{40}$K Fermi gas where  
$kT/(\hbar \omega_H)\simeq 10^2$-$10^3$ and $N\simeq 10^6$. 
By using the semiclassical approach, we have analyzed 
the spatial and momentum distributions of an ideal Fermi 
gas in harmonic potential. In particular, we have derived, 
in the large impulse approximation, the dynamic structure 
factor of the system. 
\par  
Then, we have studied the finite-temperature  
density profiles of a Fermi vapor with many hyperfine states.  
In the two-component case, by using the actual  
scattering length, we have found that  
the two components occupy the same spatial region.  
We have shown that the onset of phase-separation appears  
by increasing the scattering length or, for a fixed  
scattering length, by increasing the number of particles.  
By raising the temperature, a larger scattering length or  
a larger number of particles is needed to obtain  
the phase-separation. A Fermi vapor with three or more 
components has the same behavior but at first only one of the 
components separates from the others, which remain still mixed. 
The critical density of Fermions 
at the origin, which gives rise to the phase-separation, 
does not depend on the number of Fermi component 
and on the properties of the trap; moreover it  
satisfies the equation $n_c({\bf 0})=\pi/(48 a^3)$,  
where $a$ is the s-wave scattering length. 
\par 
Finally, we have considered the Fermi vapor is under 
rotation around a fixed axis with frequency $\Omega$.  
We have shown that, in the case of harmonic external potential,  
the rotation decreases the effective trap frequencies. 
As a consequence, the threshold for the onset of phase-separation 
occurs at larger numbers of Fermions. 
 
\newpage 
 
\section*{References} 
 
\begin{description} 
 
\item{\ [1]} Anderson M H, J.R. Ensher J R, Matthews M R, 
Wieman C E and Cornell E A 1995 {\it Science} {\bf 269} 189 
 
\item{\ [2]} Davis K B, Mewes M O, Andrews M R, van Druten N J,  
Drufee D S, Kurn D M and Ketterle W 1995 {\it Phys. Rev. Lett.} 
{\bf 75} 3969 
 
\item{\ [3]} Bradley C C, Sackett C A, Tollet J J and Hulet R G 
1995 {\it Phys. Rev. Lett.} {\bf 75} 1687 
 
\item{\ [4]} DeMarco B. and Jin D S 1999 {\it Science} {\bf 285} 1703  
 
\item{\ [5]} Holland M J, DeMarco B and Jin D S 
Preprint cond-mat/9911017.  
 
\item{\ [6]} Huang K 1987 {\it Statistical Mechanics} (New York: John Wiley) 
 
\item{\ [7]} Butts D A and Rokhsar D S 1997 
{\it Phys. Rev.} A {\bf 55} 4346  
 
\item{\ [8]} M\o lmer K 1998 {\it Phys. Rev. Lett.} {\bf 80} 1804  
 
\item{\ [9]} Bruun G M and Burnett K 1998 
{\it Phys. Rev.} A {\bf 58} 2427  
 
\item{\ [10]} Schneider J and Wallis H 1998 {\it Phys. Rev.} A {\bf 57} 1253   

\item{\ [11]} Hohenberg P C and Platzmann P M 1966 
{\it Phys. Rev.} {\bf 152} 198 

\item{\ [12]} Stenger J, Inouye S, Chikkatur A P, Stamper-Kurn D M, 
Pritchard D E and Ketterle W 1999 {\it Phys. Rev. Lett.} {\bf 82} 
4569 \\ Stamper-Kurn D M, Chikkatur A P, G\"orlitz A, 
Inouye S, Gupta S, Pritchard D E and Ketterle W 1999 
{\it Phys. Rev. Lett.} {\bf 83} 2876 

\item{\ [13]} Amoruso M, Meccoli I, Minguzzi A and Tosi M P 
Preprint cond-mat/9907370. 
 
\item{\ [14]} Pozzi B, Salasnich L, Parola A and Reatto L,  
preprint cond-mat/9911134, to be published  
in 2000 {\it J. Low Temp. Phys.} {\bf 119} N. 1/2    
 
\item{\ [15]} Inouye S, Andrews M R,  
Stenger J, Miesner H J, Stamper-Kurn D M,  
and Ketterle W 1998 {\it Nature} {\bf 392} 151 

\item{\ [16]} Stringari S 1999 {\it Phys. Rev. Lett.} {\bf 82} 4371 
 
\end{description} 
 
\newpage  
 
\begin{center} 
\begin{tabular}{|ccccc|}  
\hline \hline  
$T$ & $E/(kN)$ & $\langle\rho^2\rangle^{1/2}$  
& $\langle\rho^2\rangle^{1/2}$ & $n({\bf 0})$   \\   
\hline   
$0.05$  & $0.44$  & $9.10$  & $44.60$  & $14.12$ \\  
$0.27$  & $0.88$  & $12.89$ & $63.43$  & $7.37$ \\  
$0.55$  & $2.66$  & $17.67$ & $87.26$  & $3.15$  \\  
$0.82$  & $2.48$  & $21.56$ & $106.55$ & $1.76$  \\  
$1.10$  & $3.30$  & $24.84$ & $122.90$ & $1.15$  \\  
\hline \hline  
\end{tabular}  
\end{center}  
{\small TABLE 1. Some properties of the ideal Fermi gas  
with $N=7\cdot 10^{5}$ atoms in the harmonic trap  
($\omega_1=\omega_2=860.80$ Hz, $\omega_3=122.52$ Hz).  
Temperature $T$ and energy per particle $E/(kN)$ in units $\mu$K,  
lengths in units $\mu$m and density $n({\bf 0})$ in units  
$10^{12}$ cm$^{-3}$.} 
 
\newpage 
 
\section*{Figure Captions}  
 
{\bf Figure 1}: Comparison between exact  
(full) and semiclassical (dashed) density profiles  
for an ideal Fermi gas in an isotropic harmonic trap.  
$kT/(\hbar \omega_H)=10^{-3}$. $N$ is the number  
of Fermions. Lengths in units $a_H=(\hbar/m\omega_H)^{1/2}$  
and densities in units $a_H^{-3}$.  
	 
{\bf Figure 2}: Density profiles of the  
$^{40}$K vapor with two components in the anisotropic  
harmonic trap. $N=0.5\cdot 10^7$ atoms for each component.  
a) $T=0.1 T_F$; b) $T= T_F$; c) $T=2 T_F$ 
($T_F=1.07$ $\mu$K). Units as in Figure 1.  
 
{\bf Figure 3}: Critical number $N$ of particles for the  
phase-separation of the 2-component $^{40}$K vapor  
as a function of the temperature $T$.  
On the top: $a=157a_0$. On the bottom:  
$a=10\cdot  157 a_0$ (full diamond),  
$a=50\cdot  157 a_0$ (open triangle),  
$a=100\cdot 157 a_0$ (full circle).   
$a_0$ is the Bohr radius.  
 
{\bf Figure 4}: Density profiles of the  
$^{40}$K vapor with two components in the anisotropic  
harmonic trap at zero temperature. $N_1=10^7$ atoms.  
a) $a=1000 a_0$; b) $a=2700 a_0$ ($a_0$ is Bohr radius).  
Units as in Figure 1.  
 
{\bf Figure 5}: Density profiles of the  
$^{40}$K vapor with three components (solid, dotted and 
dashed lines) in the isotropic harmonic trap ($\omega_H=450$ Hz).  
$N=0.5\cdot 10^7$ atoms for each component.  
(a) $T=0.5 T_F$; (b) $T=2 T_F$ 
($T_F=1.07$ $\mu$K). Units as in Figure 1.  
 
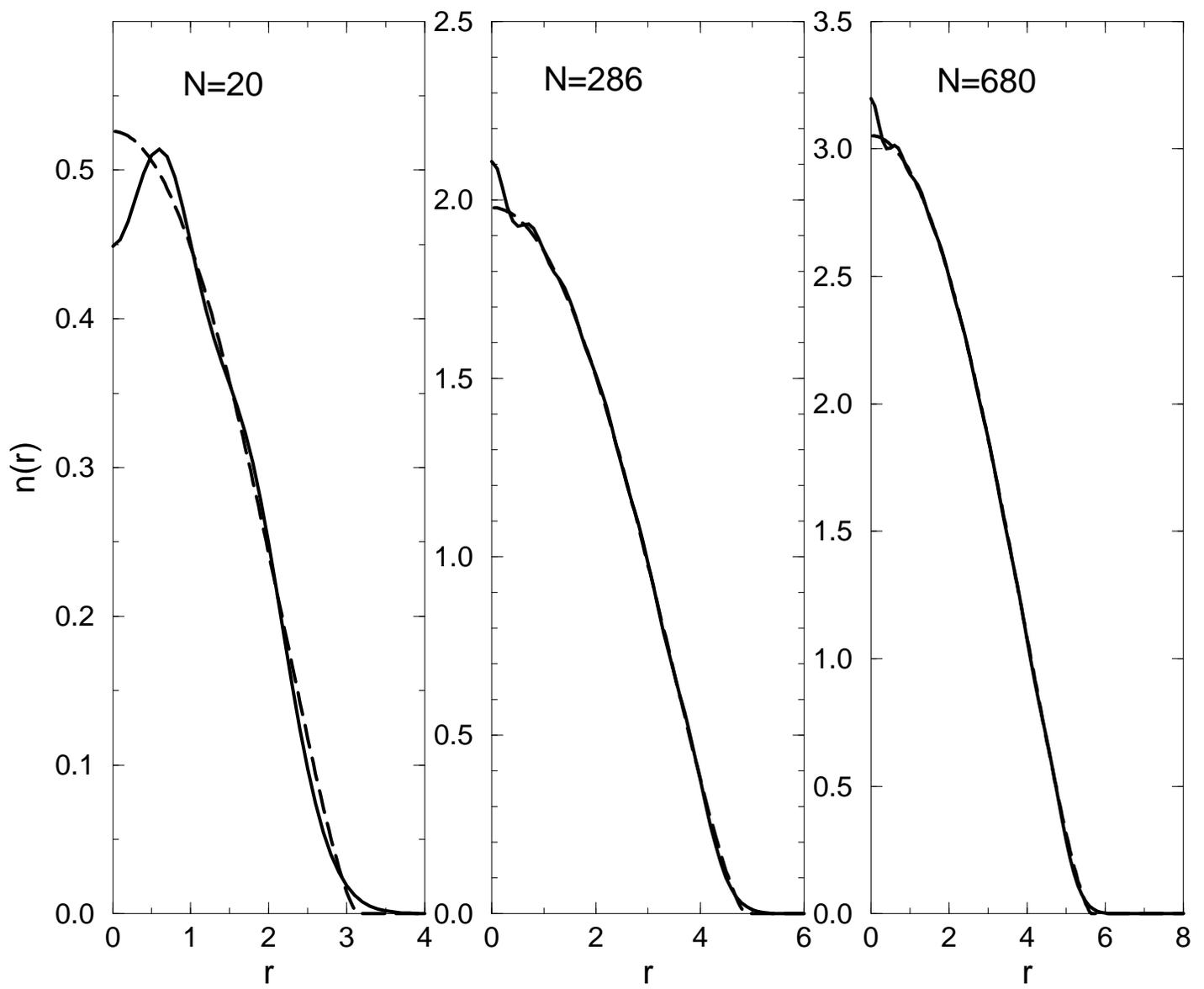
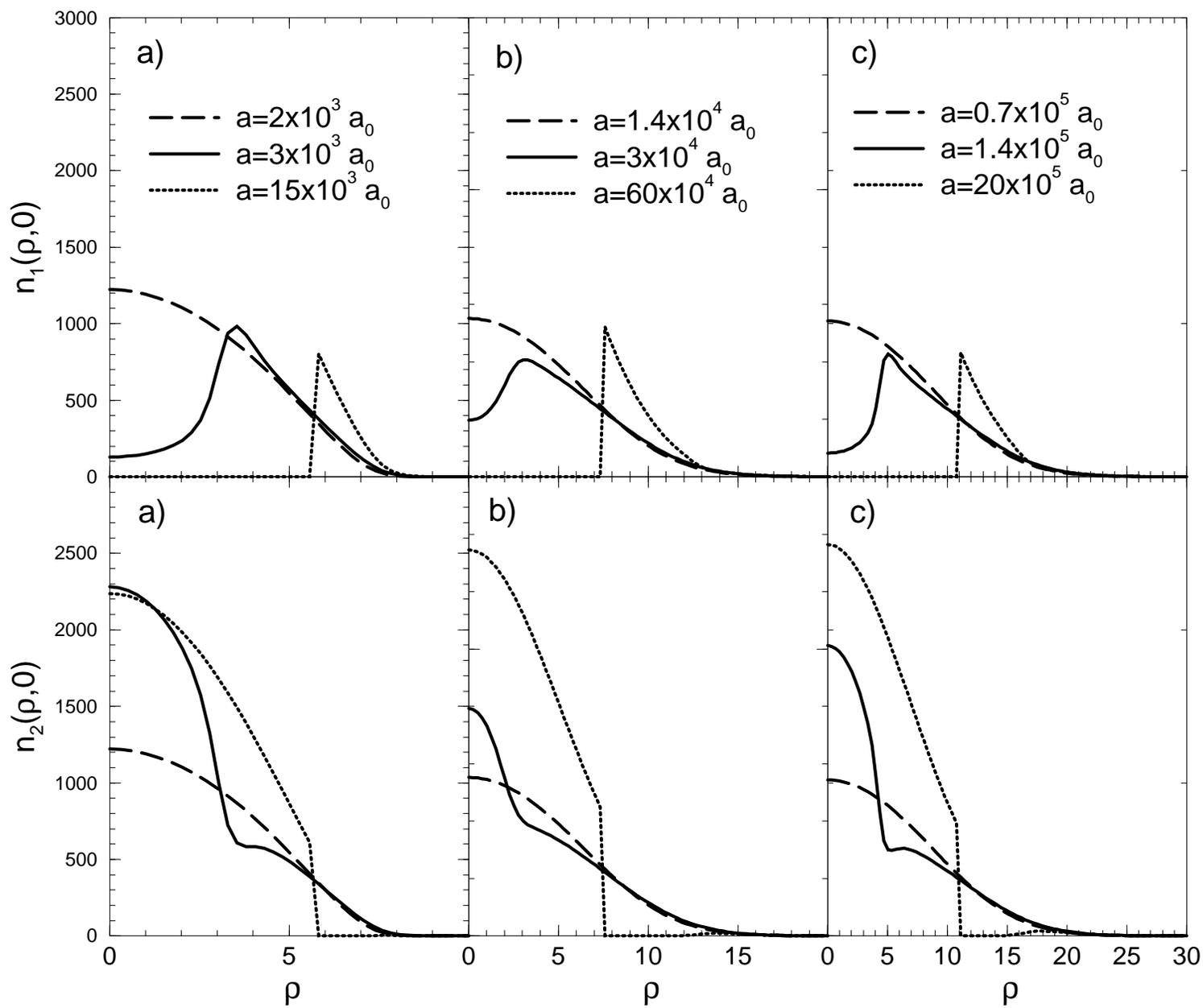
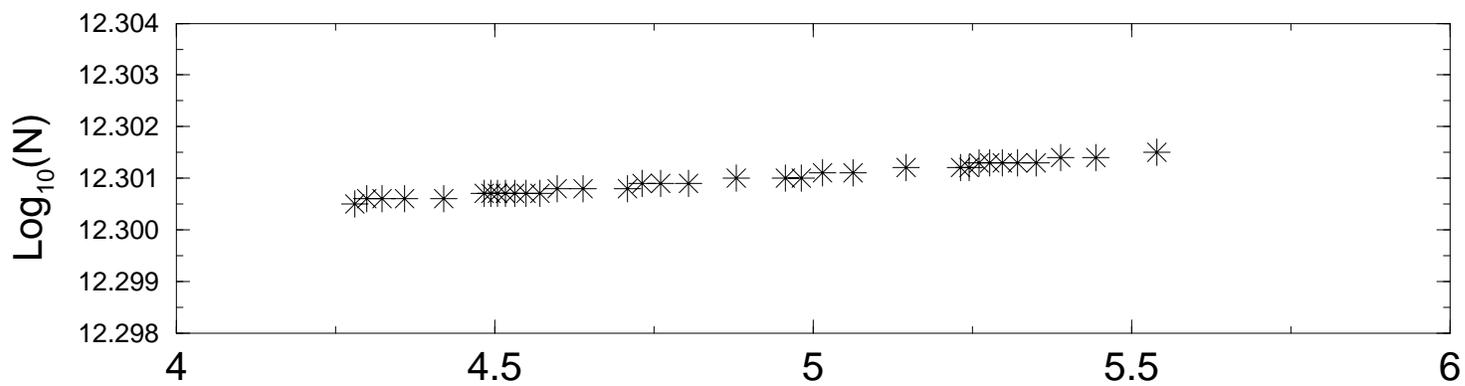
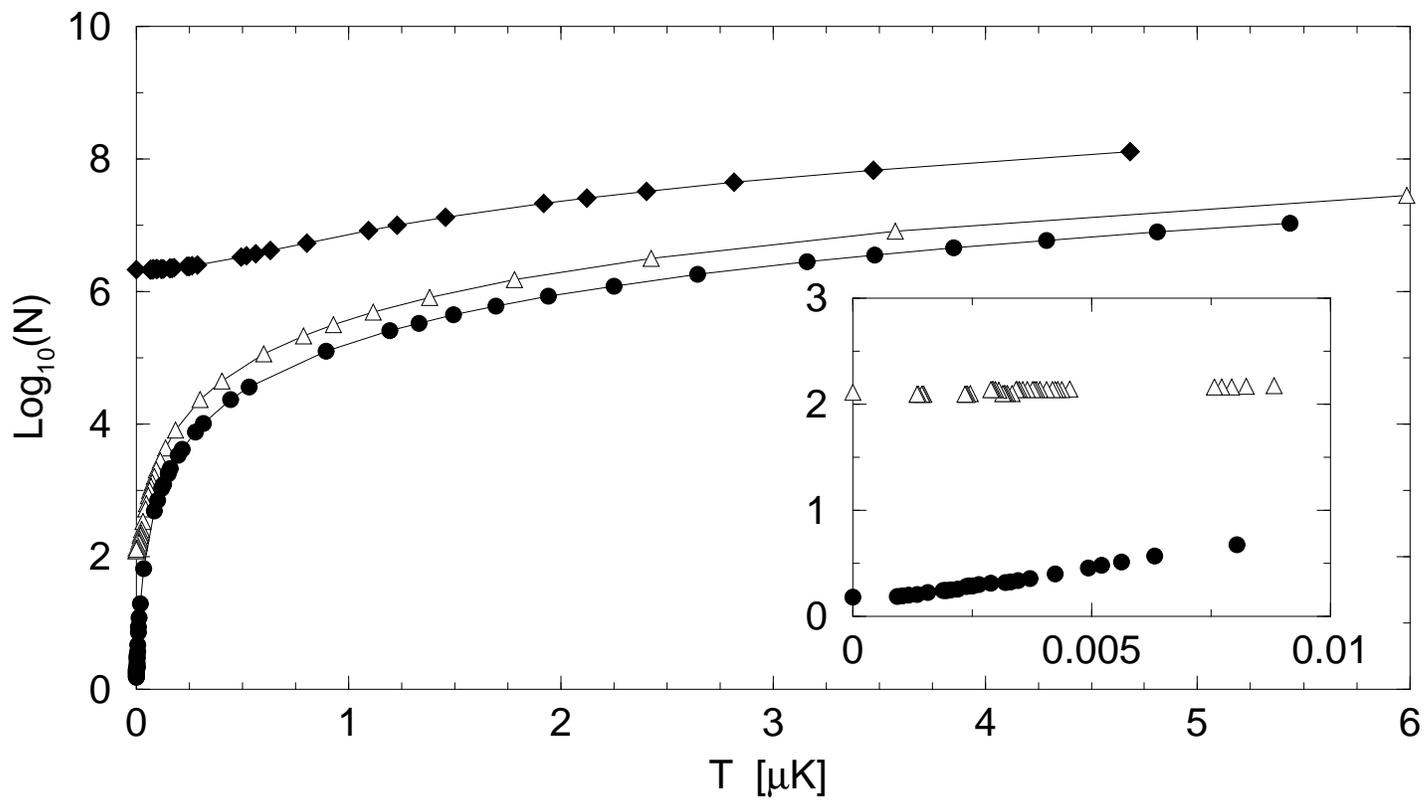
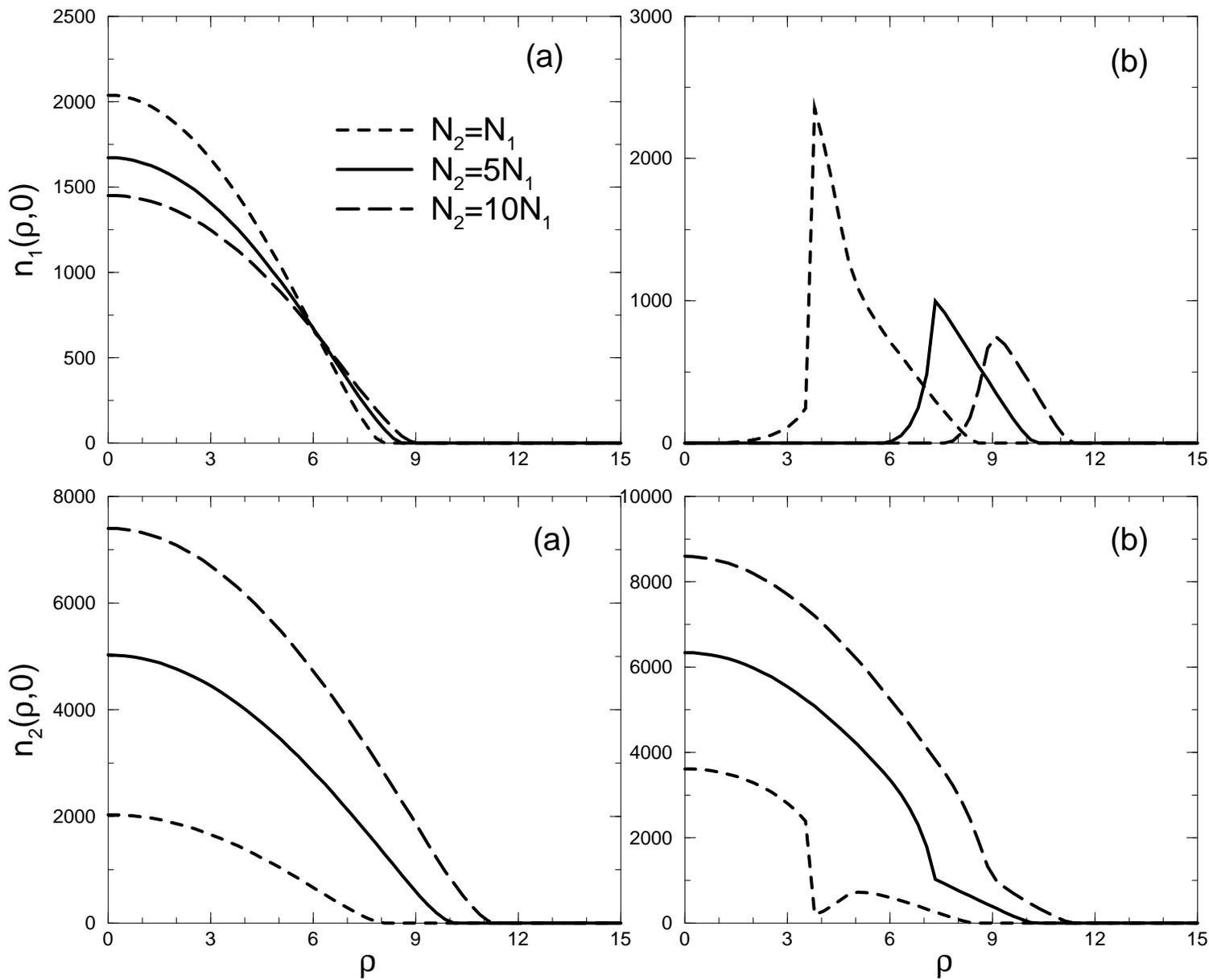
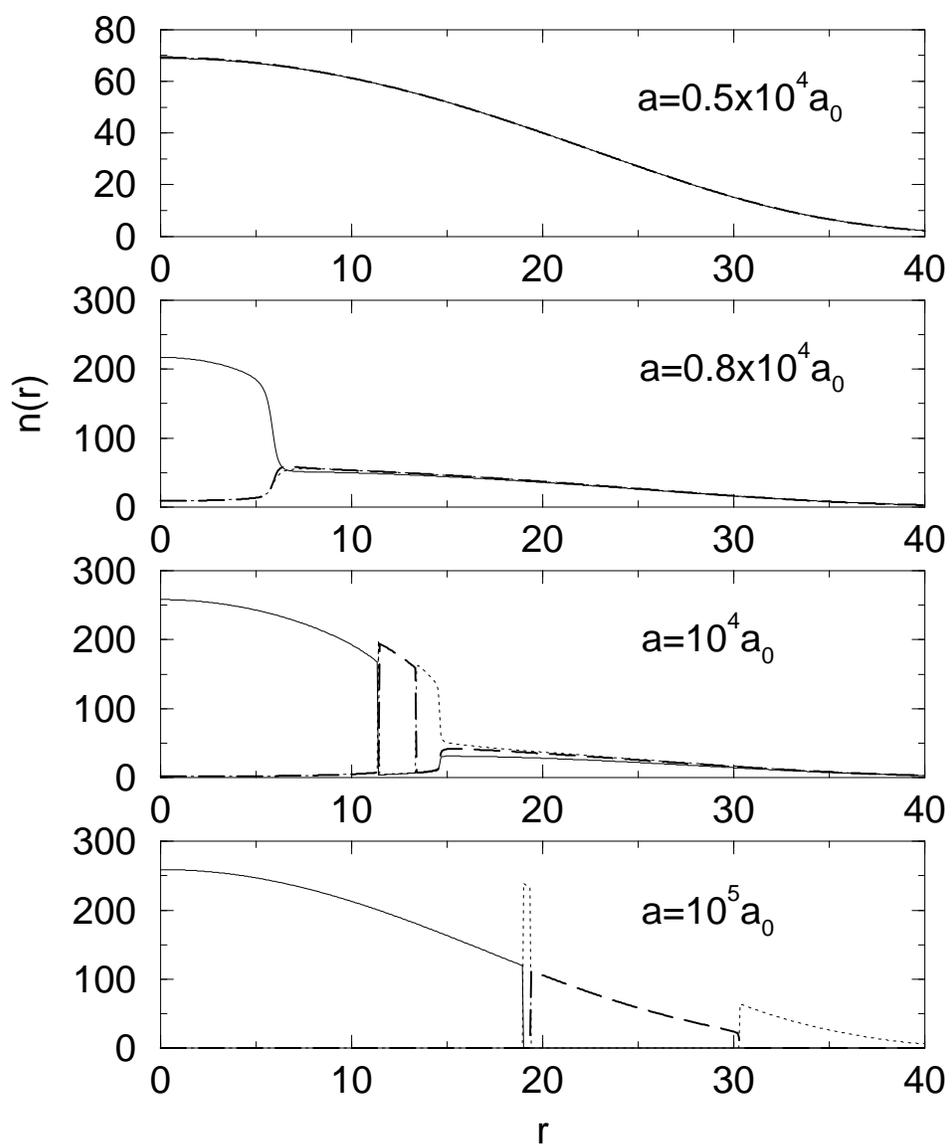
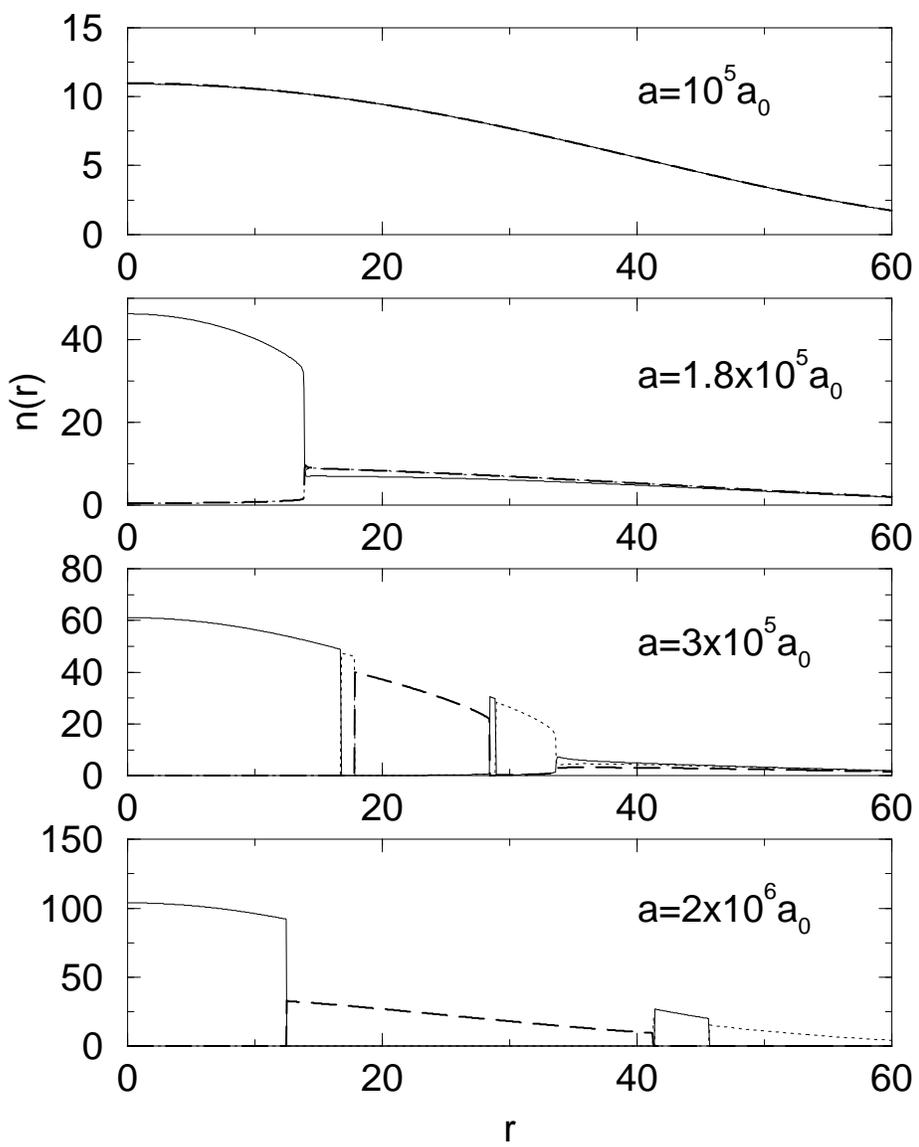
\end{document}